\documentclass[pdflatex,sn-mathphys-num]{sn-jnl}

\usepackage{graphicx}%
\usepackage{multirow}%
\usepackage{amsmath,amssymb,amsfonts}%
\usepackage{amsthm}%
\usepackage{mathrsfs}%
\usepackage[title]{appendix}%
\usepackage{xcolor}%
\usepackage{textcomp}%
\usepackage{manyfoot}%
\usepackage{booktabs}%
\usepackage{algorithm}%
\usepackage{algorithmicx}%
\usepackage{algpseudocode}%
\usepackage{listings}%

\theoremstyle{thmstyleone}%
\theoremstyle{thmstyletwo}%

\theoremstyle{thmstylethree}%

\begin{document}

\title[Article Title]{Qureka! Box - An ENSAR methodology based tool for understanding quantum computing concepts}


\author*[1,2]{\fnm{Abhishek} \sur{Purohit}}\email{abhishek.purohit@qureca.com}

\author[1,4]{\fnm{José Jorge} \sur{Christen}} \email{jorge.christen@qureca.com}

\author[3]{\fnm{Richard} \sur{Kienhoefer}}

\author[1]{\fnm{Simon} \sur{Armstrong}}

\author[1]{\fnm{Maninder} \sur{Kaur}} \email{maninder.kaur@qureca.com}

\author[1,3]{\fnm{Araceli} \sur{Venegas-Gomez}}\email{araceli.venegas-gomez@qureca.com}

\affil*[1]{\orgname{QURECA Ltd.}, \city{Glasgow}, \postcode{G2 4JR}, \state{Scotland}, \country{United Kingdom}}

\affil[2]{\orgdiv{James Watt School of Engineering, Quantum NanoPhotonics Group}, \orgname{University of Glasgow}, \city{Glasgow}, \postcode{G12 8QQ}, \state{Scotland}, \country{United Kingdom}}

\affil[3]{\orgname{QURECA SPAIN S.L.}, \city{Castelldefels}, \postcode{08860}, \state{Barcelona}, \country{Spain}}

\affil[4]{\orgdiv{Computer Engineering Department, Computer Engineering Division}, \orgname{University of Monterrey (UDEM)}, \postcode{66238}, \state{Nuevo León}, \country{Mexico}}


\abstract{As nations and organisations worldwide intensify their efforts and investments to commercialise quantum technologies and explore practical applications across various industries, there is a burgeoning demand for skilled professionals to support this rapidly growing ecosystem. With an expanding array of stakeholders from diverse professions beginning to engage with this ecosystem, there is an urgent need for innovative educational methodologies. These methodologies must not only convey the intricate principles of quantum mechanics effectively to varied professionals, enabling them to make informed decisions but also spark interest among students to delve into and pursue careers within this cutting-edge field. In response, we introduce the Experience-Name-Speak-Apply-Repeat (ENSAR) methodology, coupled with its hands-on implementation through the Qureka! Box — an innovative tool designed to demystify quantum computing for a diverse audience by emphasising a pedagogical approach rooted in experiential learning, conceptual understanding, and practical application. We present the results of deploying the ENSAR methodology using the Qureka! Box across a diverse group to validate our claims. The findings suggest a significant enhancement in the participants' grasp of foundational quantum computing concepts, thereby showcasing the potential of this approach to equip individuals from diverse professional backgrounds with the knowledge and skills to bridge the workforce demand.}




\maketitle

\section{Introduction}\label{sec1}

The emergence of quantum technologies marks a significant milestone in advancing scientific and technological capabilities by harnessing the fundamental principles of quantum mechanics for practical applications. There is a growing global effort in terms of investment, research, and commercialisation of these technologies by various stakeholders, including government organisations \cite{qureca2024}, companies \cite{Seskir2022}, and academic institutions \cite{goorney2023framework, Aiello2021} fuelled by their disruptive \cite{Moller2017, campbell2023, mosca2022} capabilities in a wide range of sectors, from pharmaceuticals to finance, defence, and more. There is a global race to harness these technological capabilities and build a quantum-ready ecosystem \cite{Purohit2024} witnessed by significant practical applications and commercialisation of the technology. 
 
 In this ever-evolving landscape of modern science and technology, quantum computing (QC) stands out as a beacon of potential, promising unprecedented computational power and capabilities that far exceed what classical systems could ever hope to achieve for certain tasks which correspond to significant scientific and daily-life problems. Yet, as we stand on the brink of the second quantum revolution, educators and learners alike confront a multitude of challenges in both teaching and understanding this nascent domain. Rooted in the complex and often counter-intuitive principles of quantum mechanics, QC demands a rethinking of foundational computational concepts, pushing the boundaries of what many have come to know and understand. The shift from binary to quantum logic, the fairly unintuitive concepts of superposition and entanglement, and the intricate mathematical underpinnings combine to create a steep learning curve. Furthermore, given the relatively recent emergence of demand to educate and train a workforce in this field, there exists a lack of standardised curricula, comprehensive teaching resources, and pedagogical strategies tailored for diverse audiences. As we venture deeper into a more quantum-ready ecosystem, it becomes imperative to address these challenges, ensuring that the next generation is adequately equipped to harness the power and potential of quantum computing.
 
In this paper, we introduce the ENSAR (Experience-Name-Speak-Apply-Repeat) methodology as a pioneering solution in a contemporary educational format, particularly for intricate concepts, such as those in quantum mechanics, facilitated by our developer educational tool, the Qureka! Box. We delve into the necessity to educate various stakeholders on the basic concepts of quantum computing, the challenges faced in the current education methodologies, and how our tool offers to solve these issues. Furthermore, we test our tool and methodology on a wide variety of professionals and students to validate our claim.

\section{Quantum Computing and Education}\label{sec2}
The advent of quantum computing (QC) is more than just another small step forward in technology; it signifies a paradigm shift in computational capabilities that has the potential to completely transform the competitive landscape of many industries. As we progress through the second quantum revolution, it is clear that QC has the potential to be one of the most disruptive technologies of the twenty-first century, similar to the transformational effects of the internet and artificial intelligence. There is a growing need for professionals with quantum skills across many industries \cite{Goorney2024, Hasanovic2022}, including healthcare, finance, logistics, and cybersecurity, according to recent studies and forecasts. These jobs are expected to grow in number significantly in the quantum sector over the next ten years. This surge is driven by the likelihood that quantum computers will perform many tasks better than classical computers, creating opportunities for novel applications and solutions that were previously thought to be computationally impractical. Furthermore, quantum technologies are projected to have significant economic impacts. Economies that can successfully harness the power of QC are likely to experience increased innovation, market competition, and the potential to generate billions of dollars in new revenue with a predicted market size of 1.3 trillion dollars for quantum technologies by 2040 and 9-93 billion dollars for solely QC \cite{mc}. Accompanying this financial growth is an anticipated creation of almost 600,000 new jobs by 2040 \cite{VenegasGomez2020}, revealing a pressing challenge: the 'quantum bottleneck' or skills shortage \cite{Kaur2022, PhysRevPhysEducRes.19.010137, Greinert2023, Amin2019}. The current quantum workforce is predominantly academic, highlighting a gap at the non-PhD level and underscoring the need for a broader educational approach. Initiatives are required not only at the tertiary level, with quantum apprenticeships and engineering programs but also at earlier educational stages to spark interest among young students \cite{faletic2023contributions, 9259951, He2021, perron2021quantum, Vishwakarma2024, laplante2022qubobsinteractiveobjectsvisual, UbbenBitzenbauer2023}. Despite various global outreach activities \cite{Seskir2022QuantumGames} aimed at raising awareness among the youth, there is an urgent call to equip teachers with the necessary knowledge and tools to educate future generations. 

Teaching and understanding QC present a plethora of challenges, stemming from the inherent complexity and abstract nature of the subject \cite{Bondani2022, Nita2023, Johnston1998}. Quantum mechanics, the foundation of QC, operates on principles that are often counter-intuitive to our daily experiences, making them difficult to grasp. This difficulty is compounded by the requirement of advanced mathematical understanding, including concepts from linear algebra and complex numbers \cite{PhysRevSTPER.11.020117}. Unlike many classical subjects, quantum phenomena often lack real-world analogies, making them challenging to relate to known experiences. Furthermore, visualising fundamental quantum concepts like superposition, entanglement, and quantum gate operations is not straightforward due to their non-classical nature.
\begin{figure}[h]
    \centering
    \begin{minipage}{\textwidth}
        \includegraphics[scale=0.2]{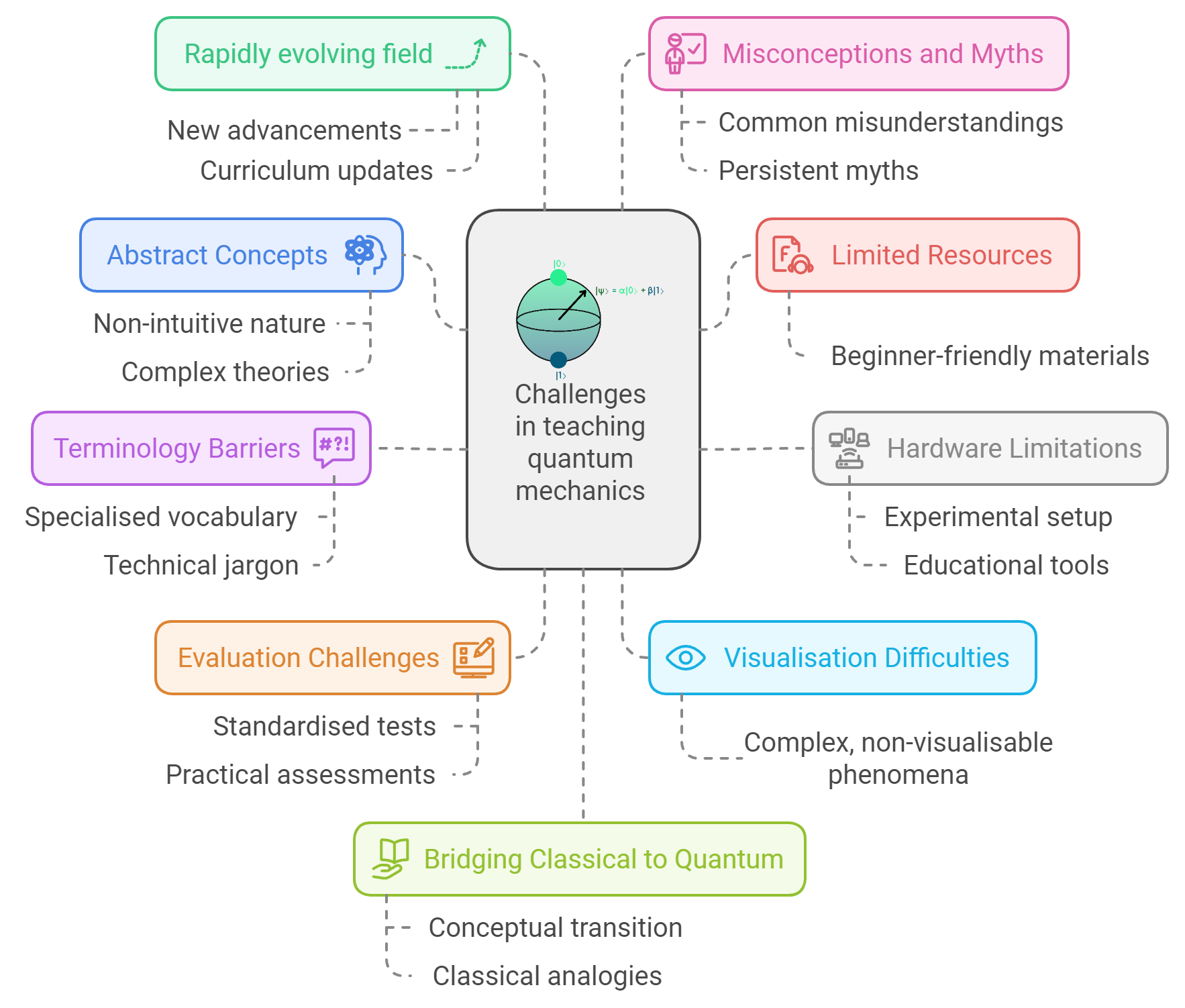}
    \end{minipage}
    \caption{Different challenges faced in teaching and understanding concepts in quantum mechanics}
\end{figure}
The field also faces a scarcity of beginner-friendly yet comprehensive educational materials, making it difficult for beginners to grasp the concepts. As a continuously evolving field, the swift advancements in quantum computation may also leave educators scrambling to keep their curriculum current. Practical demonstrations, a foundation for effective teaching in many disciplines, are hampered by the lack of easily accessible quantum computers for educational use. In addition, there are often many misconceptions \cite{Stefani2009}, some of which are fuelled by inaccurate depictions of quantum concepts in popular culture and the media. Quantum mechanics also requires a special vocabulary which can be daunting for newcomers and for those attempting to discern between closely similar notions like superposition and entanglement, the overlap of quantum concepts can make things more difficult.

Educators must decide whether to go in-depth on a few specific subjects or give a broad overview of a number of them. Transitioning students from classical to quantum paradigms is no small task and requires well-thought-out pedagogical strategies \cite{Muller2002}. Given the subject complexity, keeping students engaged and motivated throughout a quantum computation course is a constant challenge. Lastly, while the future promises numerous applications for quantum computing, the current dearth of tangible, real-world applications can make it harder for students to see the relevance of what they are learning.

\section{ENSAR Methodology}\label{sec3}

Effective teaching methodologies in quantum computation not only illuminate the underlying principles but also demystify the often counterintuitive behaviors observed in quantum systems \cite{Rafner2023, Kalkanis2003, Ke2005, 9439191}. Furthermore, as quantum computing promises revolutionary advancements in a wide range of fields, preparing learners and future innovators to harness its potential by employing pedagogical approaches that facilitate deep understanding, foster critical thinking, and nurture the capability to innovate within this disruptive technology domain becomes imperative. 

A core issue lies in the way students perceive and process scientific models. Normally, students like to give these models an amount of concreteness, viewing them not as abstract tools of representation but rather as almost literal representations of the quantum world \cite{Muller20021}. These viewpoints are in sharp contrast to those of seasoned experts who view models as tools. Reiner and Burko \cite{Reiner2003} provide additional support for this viewpoint by emphasising the crucial role thought experiments play in creating a more immersive, integrative learning environment.  As Singh \cite{Singh2009} stressed, visual aids help to further this effort by acting as vital links between the formal equations and the underlying concepts. And with the challenge of effectively replicating quantum phenomena in traditional classroom settings, the call for fresh pedagogical methodologies is clear \cite{Kersting2024}. A solid foundational understanding must be established as the potential of quantum computing grows. This offers an ideal setting for the introduction of a methodology like ENSAR, which is specifically designed to tackle these educational specifics and difficulties.

The Experience-Name-Speak-Apply-Repeat (ENSAR) methodology meticulously integrates elements from numerous recognised learning theories, establishing an innovative blend of pedagogical frameworks. Rooted in the foundational principles of experiential learning \cite{kolb, Tappert2019}, constructivism \cite{const, Pardjono2002}, active learning \cite{Partanen, TeachingLearningSTEM2016}, inquiry-based learning \cite{Rodriguez2020, Li2022, Russo2018}, and spaced repetition \cite{Tabibian2019}, ENSAR offers an integrated strategy that aims to bridge theory and exercise. ENSAR's holistic structure promises to offer educators and learners a strong framework, fostering deeper understanding and long-term retention in an educational environment where adaptability and efficiency are paramount. 

The ENSAR methodology, designed to facilitate the learning of complex concepts, is applicable across various subjects that require abstract thinking. This methodology is underpinned by three foundational principles:

\begin{enumerate}
    \item Learners do not need prior experience or knowledge to begin their educational journey.
    \item Abstract thinking abilities are not a prerequisite.
    \item Inclusivity is paramount, ensuring no participant is overlooked or left behind.
\end{enumerate}

These principles are particularly effective for diverse educational levels, including middle school, high school, and pre-graduate levels, as well as for professionals outside of engineering or science fields and general audiences.

The traditional teaching model, which has persisted for centuries, typically features a teacher presenting information in a classroom setting. This conventional format includes recognisable elements such as the use of a two-dimensional board for lectures, enforced silence with minimal peer interaction, the expectation for students to practice independently outside of class, a general prohibition against making mistakes, and a lack of differentiation in teaching methods for subjects that require abstract reasoning versus those that do not.

These traditional elements highlight the necessity for adopting the ENSAR methodology. ENSAR addresses these pedagogical gaps by promoting an inclusive, interactive learning environment that adapts to the needs of all students, particularly when introducing new and abstract concepts. By reevaluating these conventional elements, ENSAR facilitates a more engaging and effective educational experience, making it a vital approach for contemporary education in complex subjects.

\subsubsection{The Two-dimensional (2D) Board in Front of the Class}
Traditionally, concepts are conveyed using a two-dimensional surface, like a piece of paper or a board, to help recipients visualise the ideas being shared. While effective for many topics, teaching abstract, three-dimensional concepts this way often results in a significant portion of students struggling to grasp the material. The ENSAR methodology circumvents this challenge by employing 3D models that facilitate spatial-temporal understanding without necessitating high levels of abstract thinking. This transition from 3D models to 2D representations becomes a simpler, more intuitive process for students.

\subsubsection{Silence and Pay Attention}
Learning effectively involves using language, a fundamental tool in transitioning from basic knowledge acquisition to more complex understanding. Unfortunately, higher education often assumes prior knowledge of terminology, leading to confusion and disengagement. ENSAR emphasises correct naming and vocalising concepts aloud to solidify understanding. This approach leads to a classroom environment that is lively and interactive rather than silent, promoting sustainable knowledge through active participation and peer-to-peer interaction.

\subsubsection{The Student Owns, Builds, and Takes the Materials Home}
Ownership and continuous engagement with educational materials are crucial in the ENSAR methodology. Students are encouraged to take models home, allowing them to practice and share their knowledge, further embedding learning through repetition and social interaction. This method not only enhances understanding but also fosters a personal connection to the subject matter, enhancing educational outcomes.

\subsubsection{Fear of Failure and No Mistakes}
The ENSAR approach transforms the classroom atmosphere by eliminating the fear of failure. Students are reassured that mistakes are part of the learning process, referred to as iterations rather than failures. This positive reinforcement encourages continuous improvement and supports a learning environment where students feel safe and motivated. This methodology ensures that no student is left behind, fostering a collaborative and supportive learning community.

\subsubsection{All Classes are Different}
The ENSAR methodology recognises the unique needs of different subjects, especially those requiring abstract thinking. It reduces the reliance on abstract capabilities by facilitating direct interaction with 3D models before any 2D representation. This tailored approach ensures that all students, regardless of their initial understanding, can engage with and comprehend complex concepts effectively.

\subsection{The ENSAR Methodology components}
Now that we have looked at some of the most important aspects of the inhibitors for teaching abstract concepts to general students, we can take a closer look at the ENSAR methodology and its five working components, which are:
\begin{enumerate}
    \item Experience
    \item Name
    \item Speak
    \item Apply
    \item Repeat 
\end{enumerate}

In order to obtain the correct learning experience, it is imperative to apply the ENSAR components in the correct order. This means that when using the methodology, the teacher must always start by experiencing, move to naming, and then continue with the rest of the components of the methodology in order.  

\subsubsection{Experience (E)}

The ENSAR methodology begins with ``experience'' emphasising that students interact with new concepts using all senses, akin to how a child learns through tactile and sensory exploration. This stage involves the use of 3D materials or models to enhance understanding without the need for advanced abstract thinking. These models, which may include board games or card games created by the teacher, are designed to be simple yet effective in conveying complex ideas. The involvement in assembling and interacting with these 3D models ensures inclusivity and full participation, fostering a sense of ownership as students are encouraged to take these materials home to further their engagement in a comfortable setting.

\subsubsection{Name (N)}

Naming is crucial for transitioning out of what is often referred to as infantile amnesia, where early life memories are inaccessible. This is known as infantile amnesia \cite{AlberiniTravaglia2017}. We still don't have a general agreement on the main causes of this effect. However, based on the natural evidence that we build sustainable knowledge after the first 4 or 5 years, we believe that language is one of the causes that supports in moving out of this infantile amnesia effect. Looking at how children learn, initially the experience and then they learn the names of things. Parents are the first to help the children learn all the possible names. This learning process is realised most of the time by using the object in front of the child and letting them handle it while naming the object, and relating the correct name not only with the object but also with the action, car moves, a bird flies, the ball is red, etc. This stage ensures that students learn the correct terminology for new concepts, facilitating sustainable knowledge acquisition. Teachers present the names associated with the materials used, and students practice these names aloud, initially in unison and then in pairs, to reinforce their memory and understanding.

\subsubsection{Speak (S)}

Following naming, students must articulate the concepts using structured sentences, integrating the new vocabulary into their language. This active use of language in a peer-to-peer teaching format enriches learning and ensures that concepts are solidified in the students' minds. The classroom environment becomes a dynamic space for vocal expression rather than silent study, crucial for effective learning.

\subsubsection{Apply (A)}

Students are then encouraged to apply their knowledge creatively, formulating new sentences or scenarios that might even inject humor into the learning process. This not only aids in relaxing the learning environment but also helps in synthesising the new information with personal experiences, enhancing the educational impact.

\subsubsection{Repeat (R)}

The final component, ``repeat,'' underscores the importance of practice in mastering any new skill or knowledge. Repetition extends beyond the classroom, with students encouraged to engage with the material at home and share their learning with others. This continuous interaction with the educational content ensures that knowledge is not only retained but also appreciated and understood deeply.

Each component of the ENSAR methodology is designed to build on the previous steps, creating a comprehensive learning journey that makes abstract concepts accessible and enjoyable for all students, ensuring no one is left behind or aside. This methodological approach aligns with modern educational needs, especially in disciplines requiring a strong conceptual foundation such as mathematics, science, and particularly, quantum computing.

In the next section, we will apply the ENSAR methodology to develop a five-module course for the introduction to quantum computing. This course is integrated into a product named the Qureka! Box. We will describe the basic components offered in the Qureka! Box, is an educational toolkit used for assistance in learning the basic concepts of quantum computing. 

\section{Qureka! Box}\label{sec4}

Quantum computing, characterised by its reliance on mathematical formalism, presents an abstract domain that demands considerable imagination for comprehension. The Qureka! Box, applying the ENSAR methodology, serves as a pedagogical tool designed to introduce learners to quantum computing. It comprises five modules, each containing various sections that progressively build upon the user's understanding without requiring prior knowledge.

The materials within the Qureka! Box are custom-developed to facilitate an incremental learning experience through interactive, game-based approaches. This includes physical models like the Qbit Box and a 3D Bloch sphere, alongside card games that juxtapose classical and quantum computing, enhancing understanding through hands-on activities. These methods aim to demystify core quantum concepts such as superposition, entanglement, and quantum states, making them accessible and engaging. This document outlines the application of the ENSAR methodology in structuring the Qureka! Box, is aimed not just at education but at stimulating engagement through discovery and interaction. However, this document is intended as an illustrative guide rather than a comprehensive training manual. In the next sections, we will describe the conceptual design for each of the five modules included in the Qureka! Box. For each module, we include the objective, the duration of each module, and the materials used for each section in the module.

\begin{figure}[h]
    \centering
    \begin{minipage}{\textwidth}
        
        \includegraphics[scale=0.135]{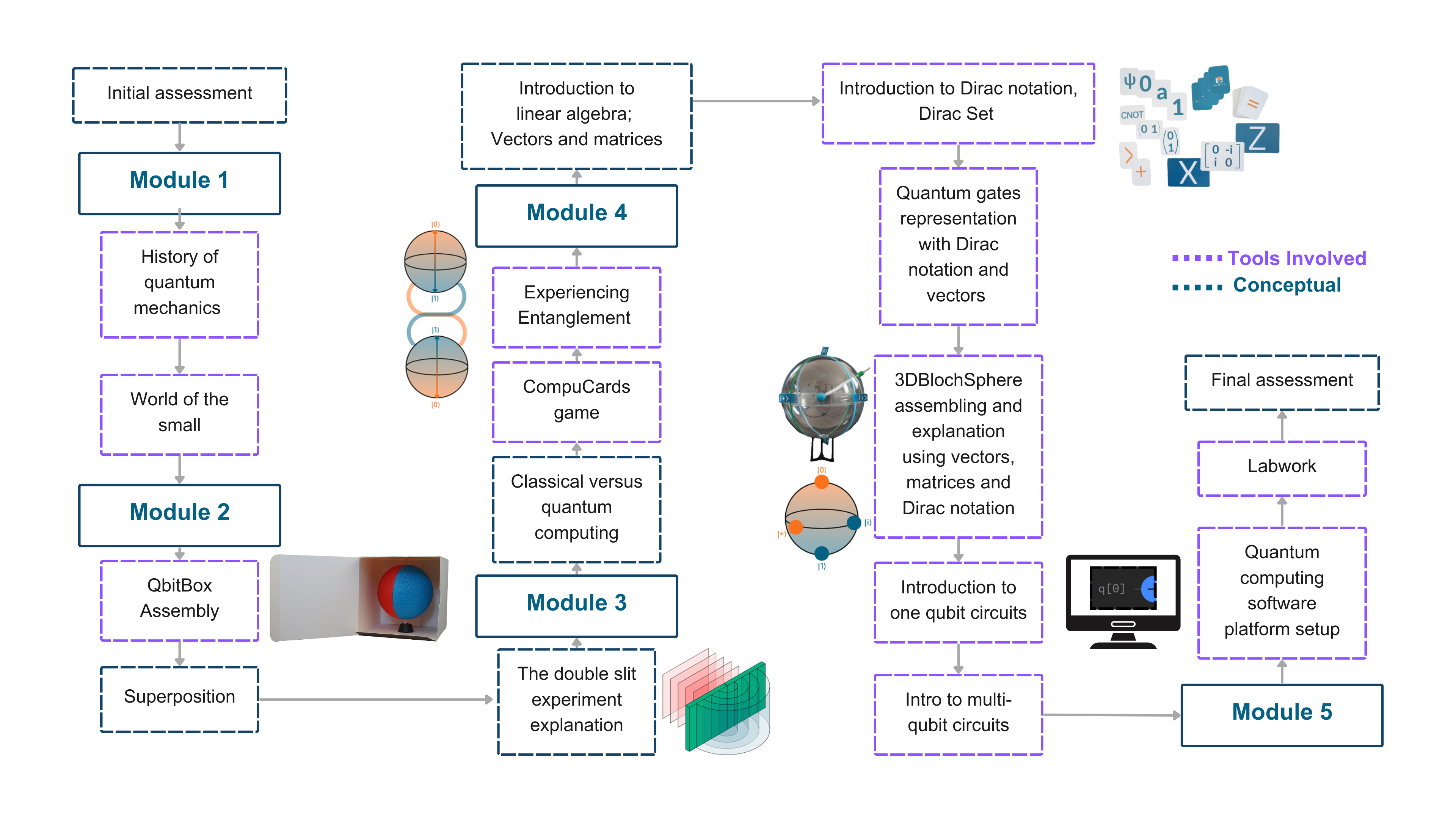}
       
    \end{minipage}
    \caption{The workflow of the Qureka! Box content, where the ENSAR methodology is applied, incorporates various interactive tools and elements to improve understanding and retention of quantum computing concepts.}
\end{figure}

\subsection{Module 1: Introduction to Quantum Computing}

The first module is designed to acquaint students with the history and fundamental concepts of the quantum world, the world of the very small. Students \textbf{experience} the timeline activity by physically placing images of key figures in quantum computing on a classroom wall. They \textbf{name} each figure and their contributions as they attach the images. This \textbf{speak} activity involves students discussing these contributions in pairs, thereby reinforcing their learning. The module employs the \textbf{apply} strategy through interactive discussions and the \textbf{repeat} technique by revisiting these figures and concepts throughout subsequent activities.

Following the historical introduction, the module explores the concept of scale in quantum mechanics through hands-on activities involving balls of varying sizes, simulating the transition from macroscopic to quantum scales. These activities help students \textbf{experience} the abstract concept of quantum scales and \textbf{name} each scale as they progress through smaller representations. This \textbf{apply} phase involves comparing the effects at different scales, and the \textbf{repeat} phase reinforces these concepts through continuous engagement with physical models.

\subsection{Module 2: Quantum Phenomena and Their Observations}

In the second module, students \textbf{experience} assembling the QbitBox, which visually and physically demonstrates quantum states and superposition. They \textbf{name} the states as they interact with the box and \textbf{speak} their observations aloud in a structured classroom discussion. The \textbf{apply} strategy is used as students manipulate the QbitBox to explore states, and the \textbf{repeat} method is evident as they repeatedly test and observe outcomes under different conditions.

The module continues with a hands-on \textbf{experience} of the double-slit experiment \cite{Young1804}, where students first \textbf{name} and \textbf{speak} about their predictions and results. They \textbf{apply} their understanding of the wave-particle duality \cite{feynman2011feynman} by adjusting variables like slit width and observing the effects, and they \textbf{repeat} these experiments to solidify their understanding through replication and discussion.

\subsection{Module 3: Classical versus Quantum Computing}

This module differentiates classical from quantum computing using the CompuCards game. Students \textbf{experience} the differences firsthand and \textbf{name} each type of computing as they engage in the simulation. They \textbf{speak} about their experiences and the advantages of quantum computing during debriefs. The \textbf{apply} phase involves analysing the outcomes of the games to draw conclusions about computational efficiency, and the \textbf{repeat} strategy is used as students engage in multiple rounds to ensure a robust understanding of the concepts.

Additionally, entanglement is explored through a magic trick that visually and conceptually demonstrates this phenomenon, enabling students to \textbf{experience} and \textbf{name} entanglement. They \textbf{apply} this knowledge through problem-solving and discussions, and \textbf{repeat} the terminology and concepts to ensure retention.

\subsection{Module 4: Visualisation and Manipulation of Quantum States}

The fourth module uses the Dirac Set and 3DBlochSphere to help students \textbf{experience} and \textbf{name} quantum states and transformations. A concise introduction to linear algebra is provided to build an intuitive understanding of the necessity for complex-number vectors and specific matrices, highlighting their parallels with quantum states in Quantum Computing. They \textbf{speak} about these states as they work through problems, \textbf{apply} their knowledge to manipulate the states using the models, and \textbf{repeat} these actions through multiple exercises to deepen their understanding and ensure proficiency.

\subsection{Module 5: Practical Quantum Computing}

In the final module, students \textbf{experience} using a real quantum computing platform, where they \textbf{name} and \textbf{speak} about the components and processes involved. They \textbf{apply} their knowledge by building and testing circuits, and they \textbf{repeat} these processes with increasing complexity to solidify their understanding and ability to predict and verify outcomes.

 By embedding ENSAR within each module, the program not only simplifies complex quantum theories and principles but also enhances student engagement and retention. This pedagogical approach, carefully tailored to demystify quantum mechanics, ensures that learners not only grasp but also apply and remember the foundational concepts. The Qureka! Box stands as a model for innovative educational tools that bridge the gap between abstract scientific theories and practical understanding, preparing students effectively for advanced studies and careers in quantum technologies.

\section{Methodology Deployment}\label{sec5}
\begin{figure}[h]
    \centering
    \begin{minipage}{\textwidth}
        
        \includegraphics[scale=0.5]{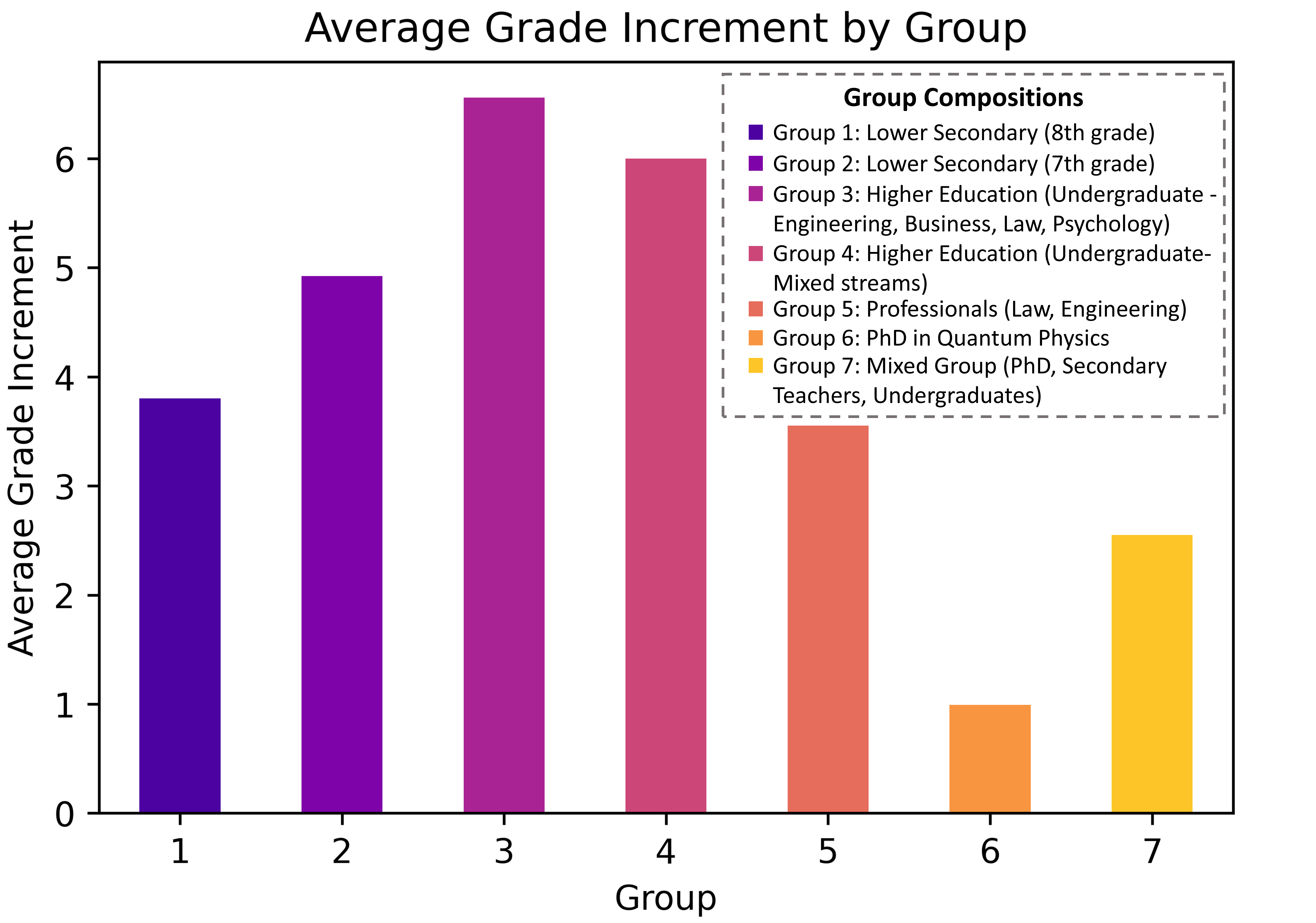}
        
    \end{minipage}
    \caption{Average grade increment across seven distinct groups, each representing different educational backgrounds. Group 1 and Group 2 consist of lower secondary education students (8th and 7th grade, respectively). Group 3 includes undergraduate students in disciplines such as Engineering, Business, Law, and Psychology. Group 4 represents undergraduates from mixed academic streams. Group 5 includes professionals in Law and Engineering. Group 6 consists of PhD candidates specializing in Quantum Physics, while Group 7 is a mixed group, comprising PhD candidates in Quantum Mechanics, secondary education teachers, and undergraduate students. The legend provides a detailed breakdown of the group compositions. Groups 3 and 4, consisting of undergraduate students, showed the highest grade improvements, validating the effectiveness of our ENSAR methodology in teaching quantum concepts to beginners across different streams.}
\end{figure}

To validate and corroborate the methodology, it was tested on seven groups of individuals with different compositions evaluated based on their performance on their initial knowledge of quantum mechanics. In our study, we employed the Qureka! Box as a tool to evaluate the effectiveness of the ENSAR methodology across a diverse array of participants. The participants were drawn from a range of distinct categories: secondary students at both junior and senior levels, university graduates from diverse disciplines such as engineering and business, individuals with backgrounds in quantum technologies, and professionals spanning various sectors. The objective was to assess the applicability and impact of the ENSAR methodology in enhancing the understanding of quantum computing concepts among individuals with different levels of prior knowledge. To establish a baseline, we started with a set of fundamental questions aimed at understanding the participants' initial understanding of quantum computing. Following their engagement with the ENSAR-based sessions, facilitated by Qureka! Box, we conducted a subsequent final assessment comprising the same set of questions as in the initial assessment, designed to measure the depth of understanding and conceptual clarity acquired by the participants. These were standard sets of questions across all groups and a comparison of initial and final grades offered valuable insights into the use of this methodology across carried groups of educational and professional backgrounds for education in quantum computing.

\begin{figure}[h]
    \centering
    \begin{minipage}{\textwidth}
        
        \includegraphics[scale=0.8]{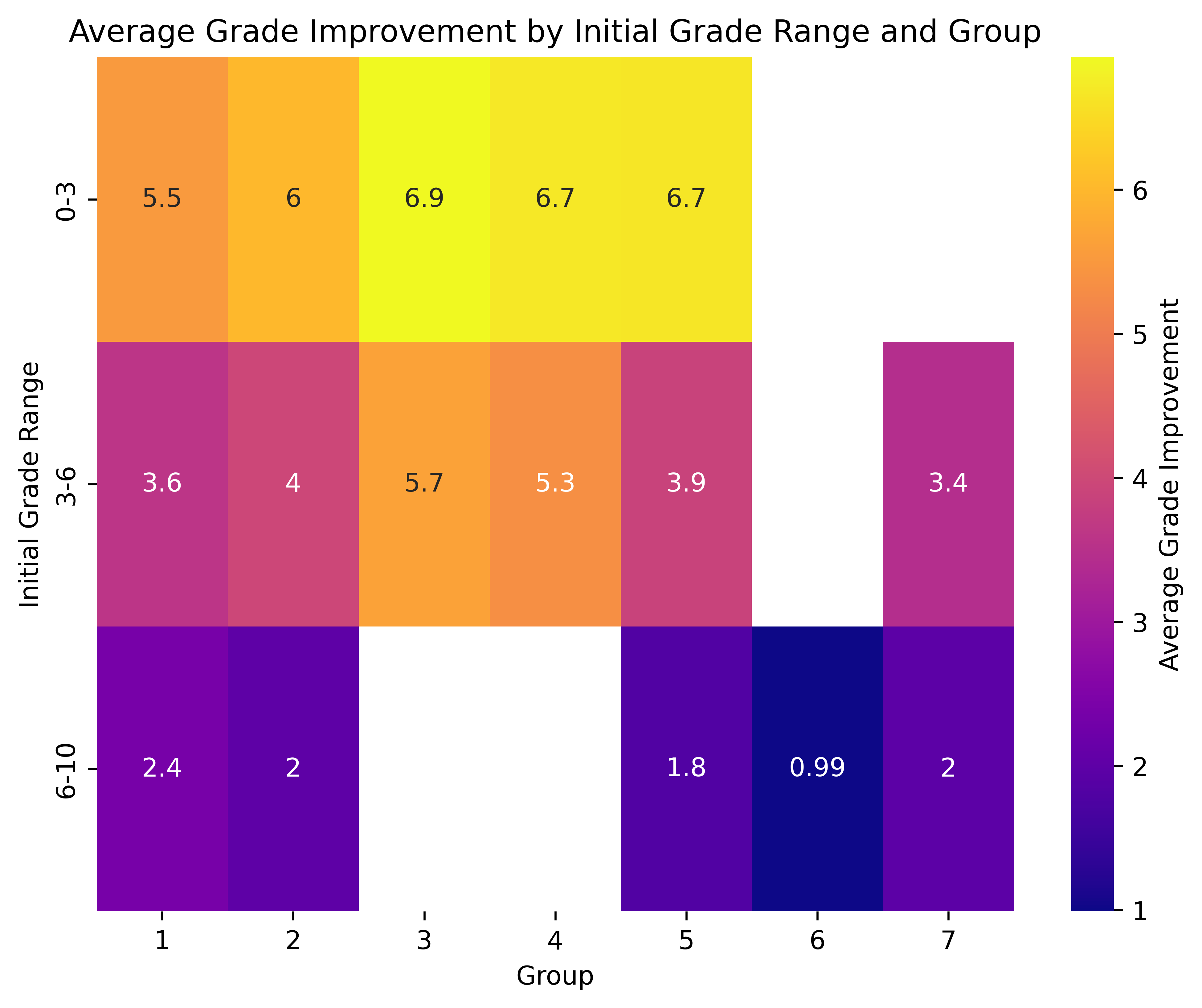}
        
    \end{minipage}
    \caption{Heat map showing an average grade increment, categorised by initial grade range. The plot shows a substantial increment of the grades for participants with little or no prior knowledge of quantum computing concepts showing the effectiveness of the tool for imparting foundational concepts in complex concepts.}
\end{figure}

The average initial grade shows an impressive 77.64{\%} average increase in grades with an average increase of 3.91 points in a total score of 10 points. The average grade improvement varies across different groups due to varied initial levels of understanding and knowledge about quantum mechanics. Group 6 which consisted of quantum technology professionals having prior education in QC and higher average grades in the initial assessment tends to have lower average grade improvements. This is expected, as those starting with higher knowledge levels have less room for improvement compared to those starting with lower initial grades. However, groups with initial grades in the 0-3 range show notable improvements, with some groups achieving average improvements greater than 5 points as seen in Fig 3 and Fig 4. Groups 3 and 4, consisting of undergraduate students from various disciplines, demonstrated the highest grade increments. Group 3, which included students from Engineering, Business, Law, and Psychology, showed the greatest improvement, followed closely by Group 4, comprised of undergraduates from mixed academic streams. This suggests that the Qureka! Box and ENSAR methodology is highly effective for individuals starting with a limited understanding of quantum physics and computation. This makes it an effective tool for understanding quantum computation for people from diverse professions and as a beginner tool in high school curriculum to introduce these complex concepts and spark their interest. This aligns with the principles of ENSAR, which aims to foster deeper understanding and long-term retention through active participation and the integration of new information with existing knowledge. The results also highlight the importance of adaptability in educational tools and methods to cater to varying levels of prior knowledge and learning paces.

\section{Conclusion}\label{sec6}

In this paper, we introduced the ENSAR methodology designed to perceive complex scientific concepts and their application to quantum computing. We present a detailed implementation of the methodology and investigation of results using the ENSAR methodology, facilitated by the Qureka! Box (our proprietary tool), providing compelling evidence of its effectiveness in addressing the educational challenges inherent in quantum technology. We present a detailed implementation of the methodology using our proprietary tool. Throughout this paper, we have detailed the pressing need for innovative educational approaches to prepare a workforce capable of navigating the complexities of quantum mechanics and computation, to cater to the bottleneck in the workforce as the field predicts exponential market and job growth and widespread application across various industries. The ENSAR methodology presented in this paper is an innovative pedagogic methodology designed for learning complex concepts and can be applied to any subject without the need of abstract thinking abilities. The implementation and deployment of the ENSAR methodology through the Qureka! Box demonstrated significant improvements in understanding quantum computing concepts further validating our methodology. In summary, when teaching abstract concepts, the ENSAR methodology tells us what to do, while the Qureka! Box shows us how to do it by implementing it in quantum computing. Our study advocates for a proactive approach in curriculum development, teacher training, and the creation of accessible educational resources to cultivate a quantum-savvy workforce. Although the ENSAR methodology is not limited to quantum computing education, we advocate the need to bridge the bottleneck in quantum workforce development and educate stakeholders from a varied background to make informed decisions to mitigate the risk of hype. As the benefits of quantum technologies are realised across the globe, fostering a future where quantum innovation is integral to solving some of society's most pressing challenges, education and advanced ways of learning are the first steps of the process.


\bibliography{sn-bibliography}

\end{document}